\documentclass[aps,prb,twocolumn,showpacs,superscriptaddress,groupedaddress]{revtex4-1}
\usepackage{amsmath}
\usepackage{amssymb}
\usepackage{soul}  
\sethlcolor{yellow}
\usepackage{url}
\usepackage{hyperref}
\usepackage[dvips]{color}
\usepackage{color}
\definecolor{bluegreen}{rgb}{0,0.2,0.8}
\usepackage{subfigure,amsmath,verbatim,moreverb}
\usepackage{multirow}
\usepackage{hhline}
\usepackage{graphicx}  
\usepackage{dcolumn}   
\usepackage{bm}        
\usepackage{amssymb}   
\usepackage{framed}
\usepackage{amsmath}

\begin{document}

\thispagestyle{empty}

\title{Role of Kohn-Sham Kinetic Energy Density in Designing Asymptotically Correct Semilocal Exchange-Correlation 
Functionals in Two Dimensions}

\author{Subrata Jana}
\altaffiliation{subrata.jana@niser.ac.in}
\affiliation{School of Physical Sciences, National Institute of Science Education and Research,
Bhubaneswar 752050, Homi Bhava National Institute, INDIA}
\author{Prasanjit Samal}
\altaffiliation{psamal@niser.ac.in}
\affiliation{School of Physical Sciences, National Institute of Science Education and Research,
Bhubaneswar 752050, Homi Bhava National Institute, INDIA}

\date{\today}

\begin{abstract}
The positive definite Kohn-Sham kinetic energy(KS-KE) density plays crucial role in designing semilocal 
meta generalized gradient approximations(meta-GGAs) for low dimensional quantum systems. It has been 
rigorously shown that near nucleus and at the asymptotic region, the KE-KS differ from its von Weizs\"{a}cker
(VW) counterpart as contributions from different orbitals (i.e., s and p orbitals) play important role. 
This has been explored using two dimensional isotropic quantum harmonic oscillator as a test case. Several 
meta-GGA ingredients with different physical behaviors are also constructed and further used to design an
accurate semilocal functionals at meta-GGA level. In the asymptotic region, a new exchange energy functional 
is constructed using the meta-GGA ingredients with formally exact properties of the enhancement factor. 
Also, it has been shown that exact asymptotic behavior of the exchange energy density and potential can 
be attained by choosing accurately the enhancement factor as a functional of meta-GGA ingredients.      
\end{abstract}

\maketitle

\section{\label{sec:intro}Introduction}
The practical applicability of theoretical aspects of modern condensed-matter systems, including electronic, 
magnetic, spectroscopic, thermodynamic and quantum chemical properties of matter have been greatly simplified 
by the advent of density functional theory(DFT)\cite{hk64,ks65}. DFT, is one of the most successful approach 
to address the complex effects of the many electron phenomenon and continues to be the same. Tremendous advances 
beyond the local density approximation(LDA) are achieved through the development of wide range of semilocal 
and non-local exchange-correlation(XC) functionals \cite{b83,jp85,pw86,b88,lyp88,br89,b3pw91,pbe96,pby96,kos,
vsxc98,hcth,pbe0,tsuneda,hse03,tpss,ae05,mO6l,pbesol,akk,revtpss,tbmbj,scan15,tm16,kaup14,lp85,adhyman78,
truhlar11,ep98,henderson08,laconstantin06,pezb98,b98,kli,tsp03} with varying properties in order to accurately 
describe the ground state in three dimensions(3D). However, cutting edge research in low dimensions including, 
atomistic to artificial structures e.g., quantum dots, modulated semiconductor layers, quantum Hall systems and 
artificial graphene has also gained momentum and keenly attracted the attention of researchers so far as the 
theoretical and experimental findings are concerned \cite{kat,rm}. The methodology applied successfully to 
describe 3D systems can not be applied directly to its two dimensional(2D) counterpart due to various limitations 
\cite{klnlhm}. Therefore, the search for accurate non-local and semilocal XC functionals for describing such 
systems is a growing research field \cite{rk,tc,amgb,hser,prhg,prvm,prg,pr1,prp,rp,sr,pr2,rpvm,prlvm,vrmp,prm,
prpg,rpp,hkprg,js,jps}. 

The higher order functionals are proposed depending on their complexity and efficiency. One rung higher of 
the 2D-LDA is 2D-GGA, which is only the functional of reduced density gradient. Despite of its grand success 
in achieving accuracy, still demands potentiality in describing various electronic regime. For example, the shell 
structure of parabolic quantum dots is visualized through the topology of electron localization factor(ELF)
\cite{pcr}. The ELF is functional of density gradient and kinetic energy density. Therefore, in order to separate 
out different physical regions using different ingredients is of prime importance. So that one region can be
distinguished from the other. In this context, the KS-KE density along with reduced density gradient play 
significant role. The KS-KE dependent functionals are knows as meta-GGAs. Not only that the meta-GGAs are the 
most accurate functional described within semilocal formalism of 2D-KS-DFT~\cite{js}. Also, the behavior of 
the KS-KE plays a significant role in designing the meta-GGA functionals. The behavior of the KS kinetic energy 
density has been studied thoroughly in 3D \cite{lcon1,lcon2,cusp1,cusp2,cusp3,cusp4,cusp5,cusp6,asy1,asy2,asy3,
asy3,asy4,asy5,asy6,asy7} but due to the aforementioned time lag between the inception of the 3D and 2D metaGGA 
XC functionals, the behavior of the KS-KE in 2D has not been explored much as compared to its 3D counterpart. 

The mainstay of this paper is to study the behavior of KS-KE in 2D by considering 2D quantum harmonic oscillator as 
a test case. It has been suggested through the ELF that ~\cite{pcr}, near nucleus (i.e. at origin) and asymptotic 
region, the KS-KE behaves one electron like and therefore reduces to von Weizs\"{a}cker KE. However, it has been
noticed that near the nuclear cusp and at the asymptotic region the KS-KE differs from its von Weizs\"{a}cker 
counterpart significantly as contributions from different orbitals play crucial role. As a matter of which, several 
2D meta-GGA ingredients can be formed by using the ratio of different kinetic energy densities because they remain 
invariant under uniform density scaling. Not only that, by constructing exactly the enhancement factor, the correct 
asymptotic behavior of the exchange energy density and potential at meta-GGA level can also be deduced. Here, we 
have considered completely a different and intriguing approach to study various regions of 2D systems. The work is 
organized as follow: In the next section we will derive a theoretical framework to thoroughly study the nature of 
KS-KE near the nuclear cusp and far away from it i.e., asymptotic region. Then, the behavior of different meta-GGA 
ingredients will be derived by taking the ratios of the kinetic energy densities at different regions of the density 
profile. Lastly, an exchange only functional from meta-GGA ingredients with correct asymptotic potential and exchange 
energy density will be constructed using formally exact properties of enhancement factor.       
             
\section{\label{sec:method}Methodology}
The total electronic energy within density functional formalism is given by,
\begin{equation}
E[\rho] = T_{KS}[\rho] + U[\rho] + \int~\rho(\vec{r})v_{ext}(\vec{r})d^2r + E_{xc}[\rho],
\label{eq1}
\end{equation}
where $\rho$ is ground state density, $T_{KS}[\rho]$ be the KS non-interacting KE, $U$ is the Hartree energy, 
$v_{ext}(\vec{r})$ is the external potential (in our case $v_{ext}(\vec{r}) = \frac{1}{2}\omega^2r^2$) and 
$E_{xc}$ is the XC energy. Only the Hartree energy and KS non-interacting KE are exactly known in terms of 
density and orbitals respectively. The exact form of XC energy is the unknown ingredient in DFT and need 
to be approximated. In DFT, the spin polarized positive definite KS kinetic energy density is defined as,
\begin{equation}
\tau^{\sigma-KS} = \sum_{i=1}^{occ}|\vec{\nabla}\Psi_{i}^{\sigma}|^2,
\label{eq2}
\end{equation} 
where $\Psi_{i}^{\sigma}$ is the occupied KS spin orbitals. But it is not unique as any term whose integral 
vanishes may be added to construct different form of the KE density. However, Eq.(\ref{eq2}) is physically 
and numerically most important because it is stable due to involvement of only the first order derivative 
of orbitals. In 3D, it has been suggested that near nucleus and asymptotic region KS-KE behaves as VW kinetic 
energy density \cite{vonw}. However, the behavior of $\tau^{KS}_{\sigma}$ for 2D has not been explored much 
which is the mainstay of this work. 

Lets now begin by considering the single electron non-interacting eigenstates of 2D isotropic harmonic 
oscillator having external potential $v_{ext}=\frac{1}{2}\omega^2r^2$, $\omega$ being the confinement 
strength of the oscillator. This simple system is very useful to gain better physical insight of the 
kinetic energy of finite systems in 2D. We will investigate in details the two important avenues i.e., 
$r\to 0$ and $r\to\infty$. The eigenstates of the above oscillator are characterized by the radial and orbital
quantum numbers $n(=0,1,2.....)$ and $l(=0,\pm 1,\pm 2....)$ and given by,
\begin{equation}
\Psi_{nl}^{\sigma}(r,\phi) = f_{nl}^{\sigma}(r)e^{-\frac{r^2}{2}}e^{il\theta},
\label{eq3}
\end{equation}
where $\sigma$(it may be $\uparrow$ or $\downarrow$) is the spin index, $r$ and $\phi$ are the radial 
distance and the azimuthal angle in the cylindrical coordinates. The density corresponding to the above
eigenstate,
\begin{equation}
\rho_{nl}^{\sigma}(r) = |\Psi_{nl}^{\sigma}(r,\phi)|^2 = [f_{nl}^{\sigma}]^2(r)e^{-r^2},
\label{eq4}
\end{equation}
where $f_{nl}^{\sigma}$ is the radial function associated with the Laguerre polynomials. The corresponding 
contribution to the KS kinetic energy density is given by,
\begin{eqnarray}
\tau^{\sigma-KS}_{nl}&=&|\nabla\Psi_{nl}^{\sigma}(r,\phi)|^2\nonumber\\
&=&\Big[\frac{df_{nl}^{\sigma}}{dr}-rf_{nl}^{\sigma}\Big]^2e^{-r^2}+\frac{l^2\rho_{nl}^{\sigma}}{r^2}~.
\label{eq5}
\end{eqnarray}
The $1^{st}$ term on the right side of Eq.(\ref{eq5}) is related to VW-kinetic energy density. The 
VW-kinetic energy density is obtained by substituting the expression of $\rho_{nl}{^\sigma}$, Eq.(\ref{eq4}) 
in VW-kinetic energy density i.e. $\tau^{\sigma-VW}=\frac{|\vec{\nabla}\rho^{\sigma}|^2}{4\rho^\sigma}$ and 
together with Eq.(\ref{eq5}) one obtains,  
\begin{equation}
\tau^{\sigma-KS}_{nl}=\tau^{\sigma-VW}[\rho_{nl}^{\sigma}] + \frac{l^2\rho_{nl}^{\sigma}(r)}{r^2}.
\label{eq6}
\end{equation}
This is the paramount equation of our investigation. Since, the total positive defined KS kinetic energy 
density is
\begin{equation}
\tau^{KS}=\sum_{\sigma=\uparrow or \downarrow}\sum_{nl}\tau^{\sigma-KS}_{nl}
\label{eq7}
\end{equation}
As, in general $\tau^{VW}\neq\sum_{nl}\tau^{VW}[\rho_{nl}^{\sigma}]$, i.e., total VW-KE density may or 
may not be equal to the sum of separate orbital VW-KE density. Therefore, the KS kinetic energy density 
in general not equal to $\sum_{nl}\tau^{\sigma-KS}_{nl}$. But equality holds only for linear case. Thus, 
Eq.(\ref{eq6}) is valid for any shell of the parabolic external potential for 2D system. Now, we will 
rigorously elaborate upon the behaviors of KS kinetic energy in two physically important regions, namely 
at $r\to 0$ and $r\to\infty$.

\subsection{\label{sec:cusp}Near origin or $r\to 0$ behavior}
Near origin or $r\to 0$ region, the density of electrons embedded inside the 2D isotropic harmonic 
oscillator is $\rho_{nl}^{\sigma}(\vec{r}\to 0) \to [f_{nl}^{\sigma}(\vec{r}\to 0)]^2 \to A_{nl}^{\sigma}r^{2l} 
(l\ge 1)$~\cite{qho}, where $A_{nl}^{\sigma}$ is associated with Laguerre polynomials. Now, Eq.(\ref{eq6}) in 
$r\to 0$ limit reduces to,
\begin{equation}
 \tau^{\sigma-KS}_{nl}(r\to 0)=\tau^{\sigma-VW}[\rho_{nl}^{\sigma}](r\to 0) + l^2A_{nl}^{\sigma}r^{2l-2}.
 \label{eqx}
\end{equation}
So the polynomial~Eq.(\ref{eqx}) has contributions only from $l=0$ and $l=1$. Polynomial terms equal or 
greater than $l=2$ vanishes as described below.\\
For $l=0$:
\begin{equation}
 \tau^{\sigma-KS}_{n0}(r\to 0)=\tau^{\sigma-VW}[\rho_{n0}^{\sigma}](r\to 0),
 \label{eqa}
\end{equation}
$l=1$:
\begin{equation}
 \tau^{\sigma-KS}_{n1}(r\to 0)=\tau^{\sigma-VW}[\rho_{n1}^{\sigma}](r\to 0) + A_{n1}^{\sigma}
 \label{eqb}
\end{equation}
and $l=2$:
\begin{equation}
 \tau^{\sigma-KS}_{n2}(r\to 0)=\tau^{\sigma-VW}[\rho_{n2}^{\sigma}](r\to 0).
 \label{eqc}
\end{equation}
As $r\to 0$, $\tau^{\sigma-VW}[\rho_{n2}^{\sigma}](r\to 0)=0$ implies the right side of Eq.(\ref{eqc}) vanishes. 
Thus, all terms including $l\geq2$ vanishes in limit $r\to 0$ and only survival terms are $l=0$ and $l=1$.
In this region, the density $\rho_{nl}^{\sigma} =[f_{nl}^{\sigma}]^2$ for $l\geq 1$ and $[f_{nl}^{\sigma}]^2
=A_{nl}^{\sigma}r^{2l}$, where $A_{nl}^{\sigma}$ is associated with Laguerre polynomials. From Eq.(\ref{eq6}),
it is clearly evident that all terms including $l\geq2$ vanishes in limit $r\to 0$ only survival term is $l=1$. 
Now upon using the density expression in $l=1$ term,
\begin{equation}
\tau^{\sigma-VW}[\rho_{n1}](r\to 0) = \frac{1}{4}\frac{(2A_{n1}^{\sigma}r)^2}{A_{n1}^{\sigma}r^2}=A_{n1}^{\sigma}~.
\label{eq8}
\end{equation}
For $l=0$, it depends on the external potential. So we have the following sequences,
\begin{eqnarray}
\tau^{\sigma-KS}_{nl}(0)=
\Bigg\{\begin{tabular}{ccc}
$\tau^{\sigma-VW}[\rho_{n0}^{\sigma}](0)$~~~~~~~~~~~~~~~~~~~~~$l=0$\\
$A_{n1}^{\sigma} + A_{n1}^{\sigma} = 2\tau^{\sigma-VW}[\rho_{n1}^{\sigma}](0)$~$l=1$\\
$0$~~~~~~~~~~~~~~~~~~~~~~~~~~~~~~~~~~~~~~$l\geq2$ 
\end{tabular}
\label{eq9}
\end{eqnarray} 
The achievement from Eq.(\ref{eq9}) is that it is linear in $\rho_{n1}^{\sigma}(0)$ and $A_{nl}$, i.e,
\begin{eqnarray}
\tau^{\sigma-VW}[\rho_{l=0}^{\sigma}]&=&\sum_{n}\tau^{\sigma-VW}[\rho_{n0}^{\sigma}](0)\nonumber\\
\tau^{\sigma-VW}[\rho_{l=1}^{\sigma}]&=&\sum_{n}\tau^{\sigma-VW}[\rho_{n1}^{\sigma}](0)~.
\label{eq10}
\end{eqnarray}
Here, summation over principal quantum number $n$ extract all the $l=0$  and $l=1$ shell contributions. 
This linear superposition is valid only at $r\to 0$ limit. Now, Eq.(\ref{eq9}) together with Eq.(\ref{eq10}) 
results in,
\begin{eqnarray}
\tau^{KS}(0)&=&\sum_{\sigma}\sum_{nl}\tau^{\sigma-KS}_{nl}(0)\nonumber\\
&=&\sum_{\sigma}\sum_n\Big[\tau^{\sigma-VW}[\rho_{n0}^{\sigma}](0)+\tau^{\sigma-VW}[\rho_{n1}^{\sigma}](0)\Big],
\label{eq11}
\end{eqnarray}
where
\begin{equation}
\rho_{l=0}^{\sigma}=\sum_{n}\rho_{n0}
\label{eq12}
\end{equation}  
and
\begin{equation}
\rho_{l=1}^{\sigma}=\sum_{n}\rho_{n1}.
\label{eq13}
\end{equation}
The $l$ value collects the contribution for each $n$ value, where $n$ runs up to maximum radial quantum number 
starting from $0$ or $1$. Hence, not only $l=0$ but also $l=1$ orbital plays significant role in the total
VW-KE density at $\vec{r}\to 0$.

\subsection{\label{sec:asymp}Asymptotic or $r\to\infty$ behavior}
Asymptotic region is equally important in density functional formalism as several exchange energy functionals 
have been designed using the asymptotic behavior of exchange potential, $v_x$ or energy density, $\epsilon_x$
such as 2D-B88 \cite{vrmp}, 2D-BR \cite{prhg} and 2D-BJ \cite{prp} functionals. However, the property of KS-KE 
in this region is very useful in designing functionals for 2D systems with correct asymptotic behavior at 
meta-GGA level. Construction of meta-GGA type functional for 2D systems recently gained momentum because of 
first ever construction of a density matrix expansion based exchange energy functional \cite{js}. In asymptotic 
region, the contributions of the outermost valence shells only count. We denote the outer shells as $n\to n'$, 
$l\to l'$ and $\rho^\sigma \to \rho'^\sigma$, where the prime denotes the asymptotic quantities. Now we consider, 
Eq.(\ref{eq6}) and from this we define a quantity $\tau'^{\sigma}$ which accounts for the deviation of the KS 
kinetic energy density from its VW counterpart i.e.,
\begin{equation}
\tau'^{\sigma}=\tau^{\sigma-KS}-\tau^{\sigma-VW}\longrightarrow_{r\to \infty}\tau^{\sigma-KS}_{n'l'}-
\tau^{\sigma-VW}_{n'l'}=\frac{l'^2\rho^{'\sigma}_{n'l'}}{r^2}.
\label{eq14}
\end{equation}
From the above expression, it is quite clear that if we consider $l'=0$ type outer shell, then $\tau^{\sigma-KS}$ 
exactly approaches to $\tau^{\sigma-VW}$. On the other hand if $l'\neq 0$, then the contributions of the other 
outer shells also count. In that case, $\tau^{'\sigma}$ decays as $\sim\frac{1}{r^2}$.     
\begin{figure}
\begin{center}
\includegraphics[width=3.4in,height=2.0in,angle=0.0]{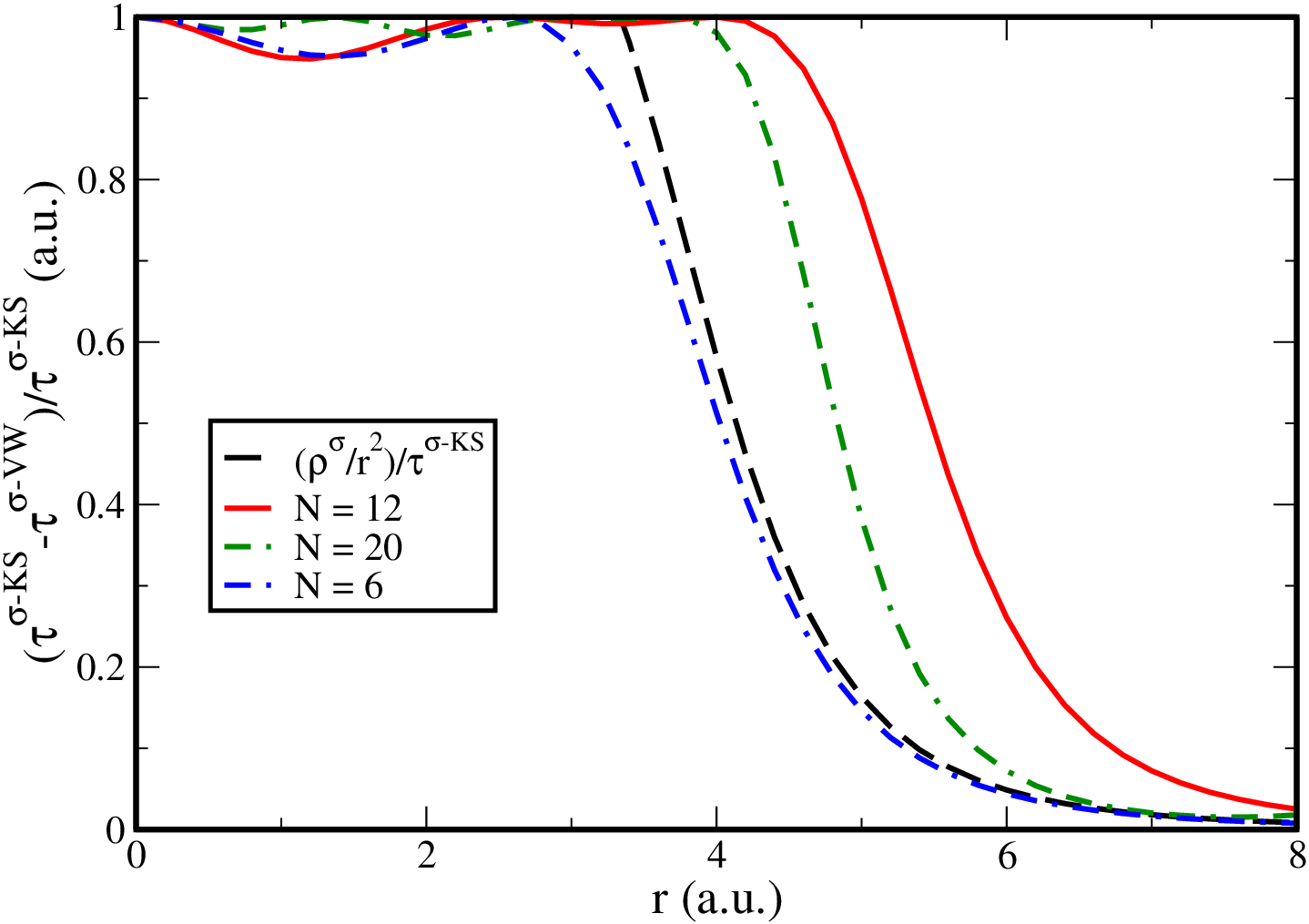} 
\end{center}
\caption{Relative deviation of the exact KS kinetic energy from the VW kinetic energy density for 
N = 6 ($\omega=0.25$), N = 12 ($\omega=1/1.89^2$) and N = 20 ($\omega=0.50$). For N = 6 ($\omega=0.25$) 
the behavior of $\frac{\rho_\sigma(r)/r^2}{\tau^{KS}_{\sigma}}$ is also shown.}
\label{fig1}
\end{figure}
In Fig.(\ref{fig1}), we have plotted the deviation of KS kinetic energy term from VW-kinetic energy density for 
$n=6, 12$ and $20$ electrons confined in a parabolic quantum dot with different confinement strengths. The behavior of 
$\frac{\rho^\sigma(r)}{r^2}\frac{1}{\tau^{\sigma-KS}}$ for $N=6$ is also shown in the figure. It is also evident that, 
as $r\to\infty$, $\tau^{\sigma-KS}\to\tau^{\sigma-VW}$. In that case, $\frac{\rho^\sigma(r)}{r^2}\to 0$. Thus, in 
$r\to\infty$, the correct VW behavior is achieved by KS kinetic energy density. Here, all the calculations are done 
using KLI \cite{kli} exact exchange(EXX) as implemented in OCTOPUS \cite{octopus} code and the output is used as the 
reference input for our calculations. 
  
\section{\label{sec:ing}Behavior of meta-GGA ingredients}
\begin{figure*}
\centering
  \begin{tabular}{@{}cccc@{}}
    \includegraphics[width=3.4in,height=2.0in,angle=0.0]{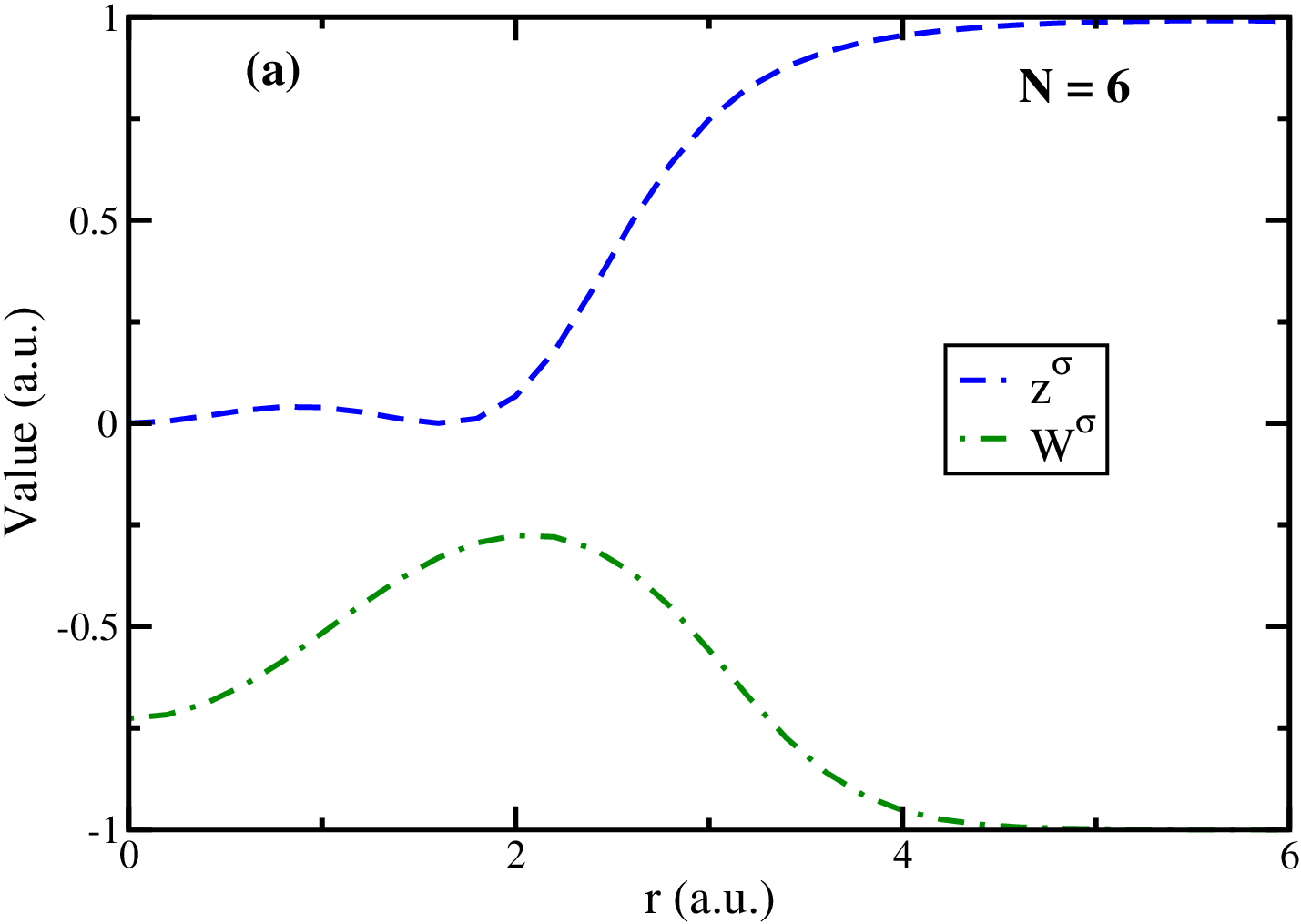} &
    \includegraphics[width=3.4in,height=2.0in,angle=0.0]{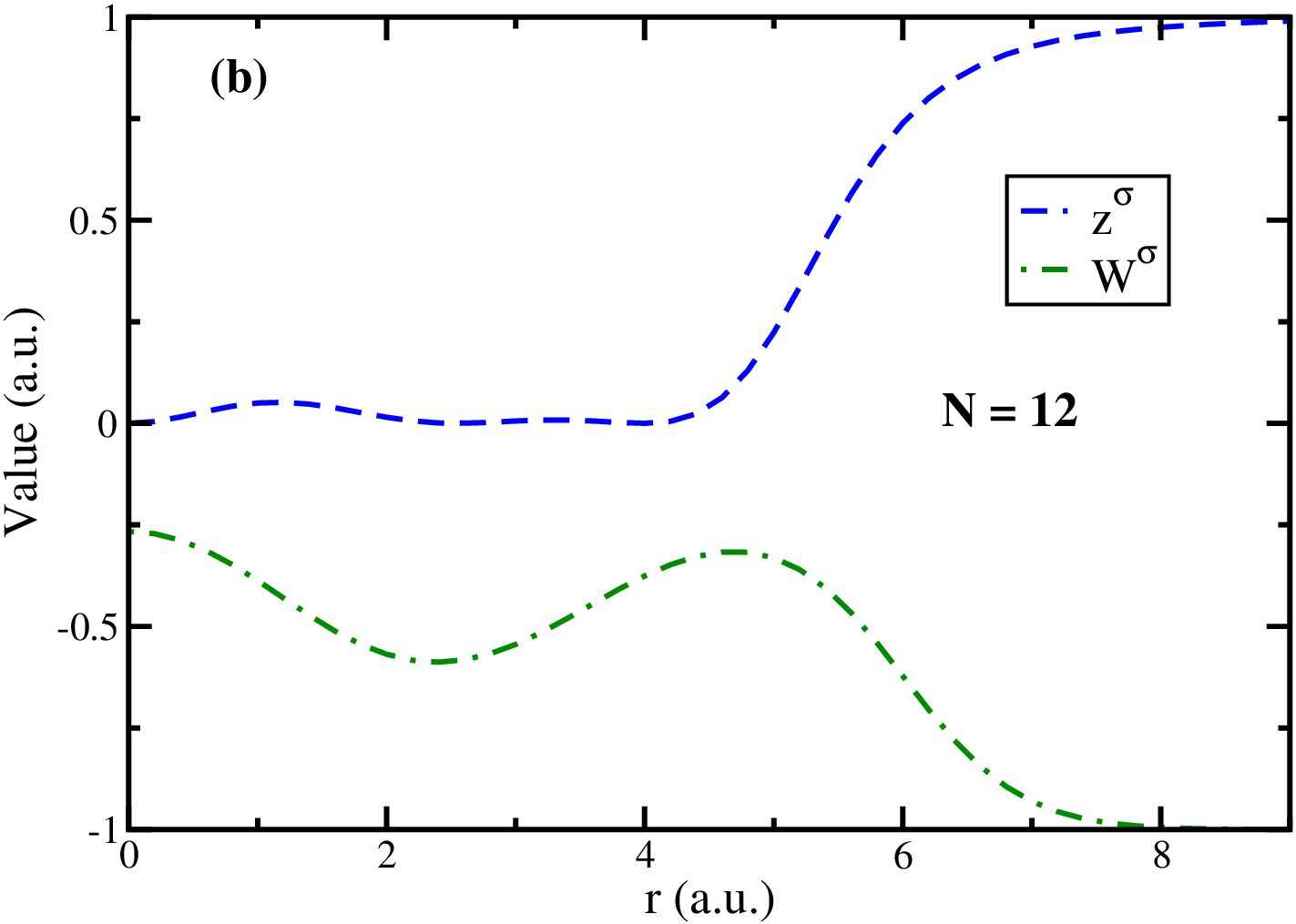} \\
    \includegraphics[width=3.4in,height=2.0in,angle=0.0]{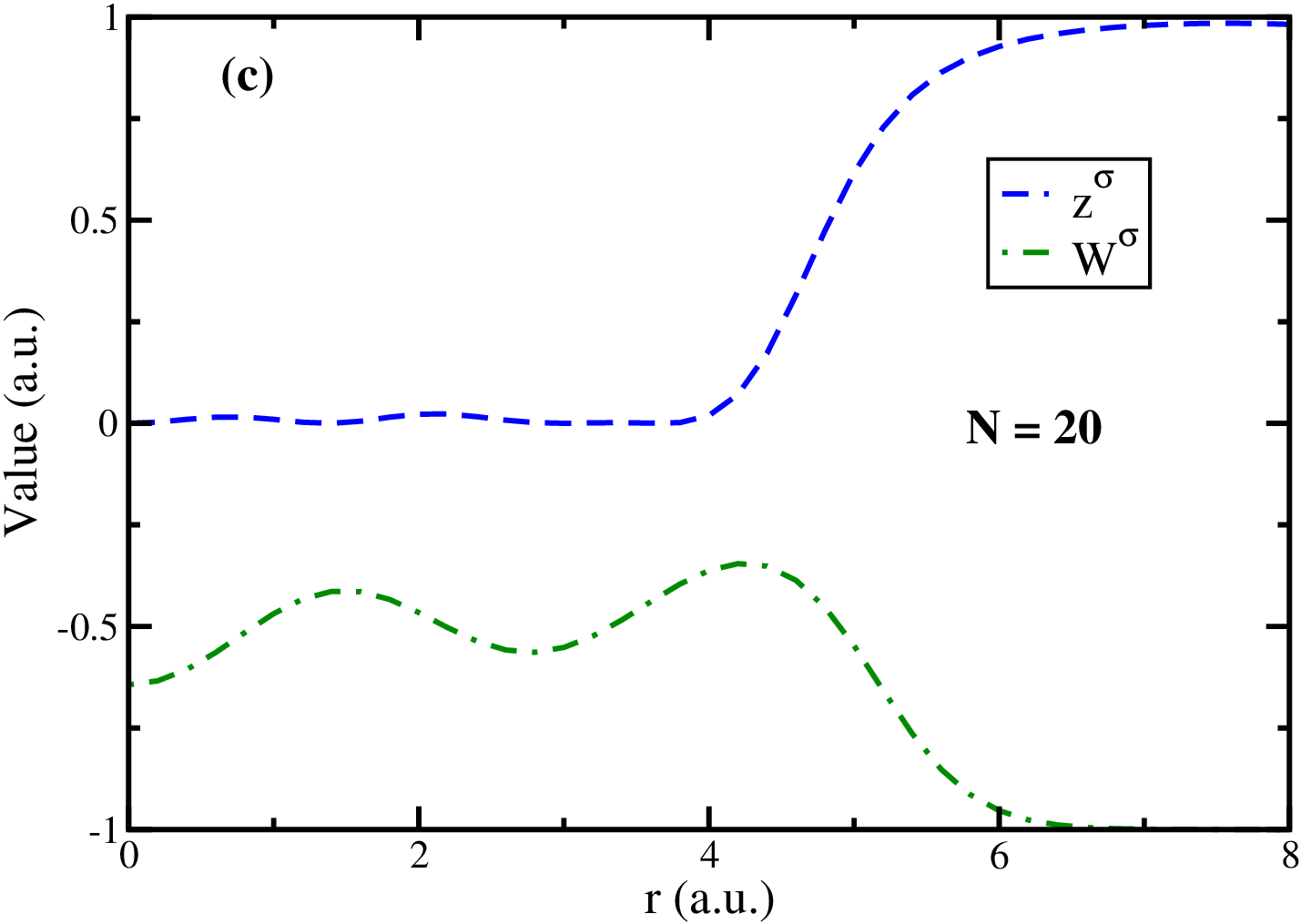} &
    \includegraphics[width=3.4in,height=2.0in,angle=0.0]{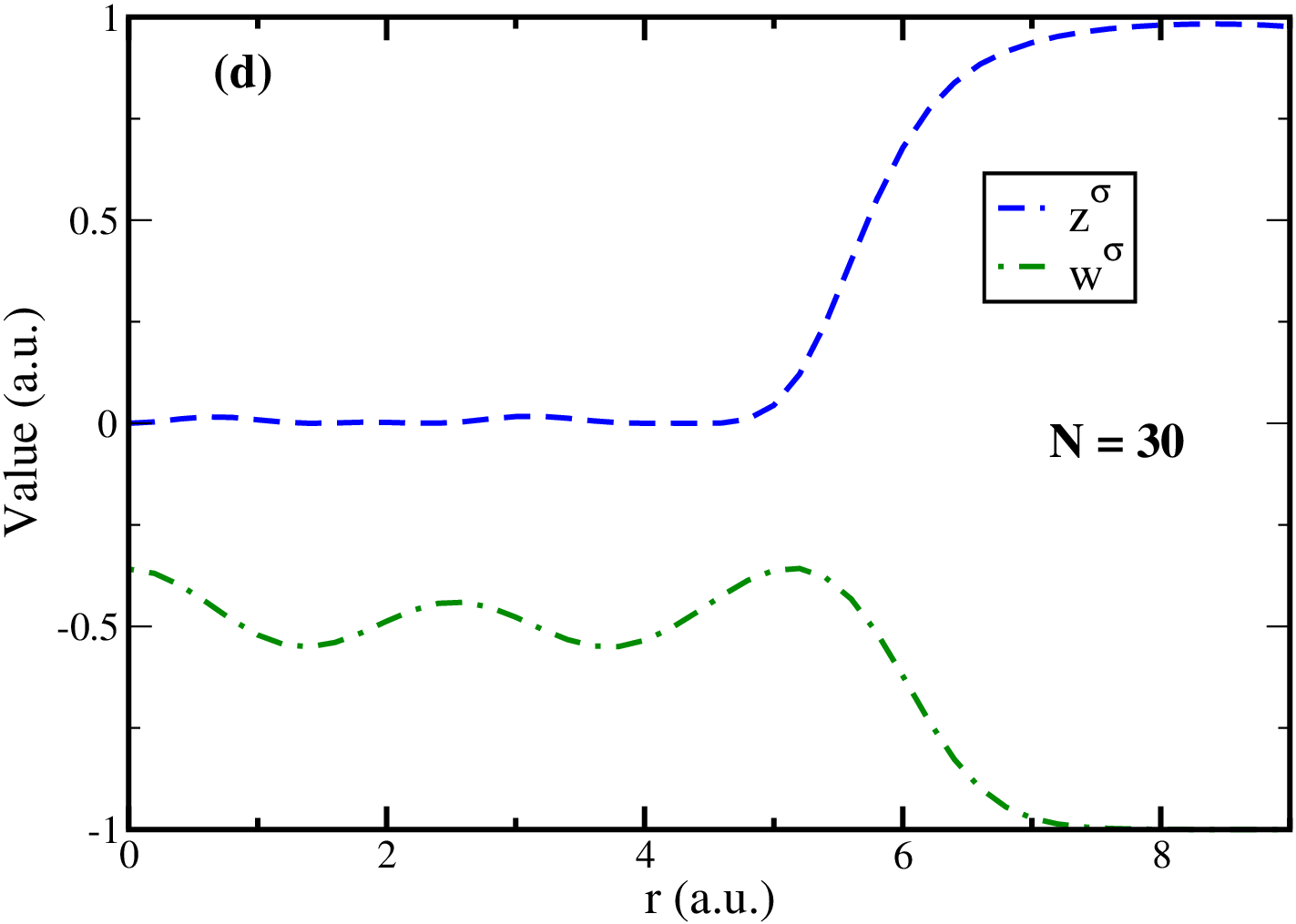}
  \end{tabular}
  \caption{The $z^{\sigma-KS}$ and $W^{\sigma-KS}$ meta-GGA ingredients for $N-$particles $(=6, 12, 
20, 30)$ in a parabolic quantum dots with varying confinement strength.}
\label{fig2}
\end{figure*}
The nature of the KS-KE (as discussed in previous section), is important to design different meta-GGA ingredients 
depending on their behavior in different regions for finite systems in 2D. In 3D, the KS-KE, VW and uniform 
kinetic energy density are used to design several reduced density gradients for meta-GGA functionals. Analogously,
in 2D, different meta-GGA ingredients can also be constructed. In 2D, KS and VW kinetic energy densities remains 
invariant under uniform density scaling i.e. $\tau^{\sigma-X}[\rho_\gamma] = \gamma^4\tau^{\sigma-X}[\rho(\gamma
\vec{r})]$, where X = KS~or~VW. Therefore, ratio of any two KE densities can form the ingredient for 2D meta-GGAs. 
For one electron ground state, $\Gamma^\sigma(\vec{r},\vec{r}') = \rho^{\sigma~1/2}(\vec{r})\rho^{\sigma~1/2}
(\vec{r}')$, with $\Gamma^\sigma(\vec{r},\vec{r}')$ be the reduced density matrix. Now, define a quantity
\begin{equation}
\zeta^{\sigma}=\frac{\rho^{\sigma~1/2}(\vec{r})\rho^{\sigma~1/2}(\vec{r'})}{\Gamma^\sigma(\vec{r},\vec{r}')},
\label{eq15}
\end{equation} 
which can serve as an iso-orbital indicator and equals to $1$ when $\vec{r}=\vec{r}'$. However, the quantity 
\begin{equation}
z^\sigma = lim_{r\to r'}\frac{\vec{\nabla}_r.\vec{\nabla}_r'\rho^{\sigma~1/2}(\vec{r})\rho^{\sigma~1/2}
(\vec{r}')}{\vec{\nabla}_r.\vec{\nabla}_r'\Gamma^\sigma(\vec{r},\vec{r}')}=\frac{\tau^{\sigma-VW}}
{\tau^{\sigma-KS}}
\label{eq16}
\end{equation} 
is an iso-orbital indicator, becomes $1$ for one electron or two electron singlet state (characteristic as 
$r\to 0$ and $r\to\infty$ i.e., near origin or asymptotic region) and other regions having values between 
$0$ and $1$. Also, it remains invariant under unitary transformation of KS occupied orbitals. All these 
conditions prefer $z^\sigma$ to be an adequate iso-orbital indicator for 2D but it suffers from the order 
of limit problem. To elaborate on it, we define another meta-GGA ingredient, $\alpha^\sigma=\frac{\tau^{\sigma-KS}
-\tau^{\sigma-VW}}{\tau^{\sigma-unif}}$ with ${\tau^{\sigma-unif}}=2\pi{\rho^\sigma}^2$, the uniform KE density 
in 2D. Now expressing $z^\sigma$ in terms of $\alpha^\sigma$ and reduced density gradient $p^\sigma={s^\sigma}^2
= \frac{\tau^{\sigma-VW}}{\tau^{\sigma-unif}}$ as,
\begin{equation}
z^\sigma = \frac{1}{1+\frac{\alpha^\sigma}{p^\sigma}}
\label{eq17}
\end{equation}
we obtain,
\begin{equation}
\lim_{p^\sigma\to 0}\lim_{\alpha^\sigma\to 0} z^\sigma = 1 ~\textit{but} ~\lim_{\alpha^\sigma\to 0}
\lim_{p^\sigma\to 0} z^\sigma = 0~.
\label{eq18}
\end{equation}
This problem arises in iso-orbital regions, where $p^\sigma$ approaches zero. The order of limit problem is not 
new in 3D \cite{ptss}. Exemplification of this problem for 2D finite system is shown in Fig.(\ref{fig2}). The 
meta-GGA ingredient, $\alpha^\sigma$ is $0$ for one or two electron singlet state and becomes $1$ in uniform density 
limit. This can also be used to form the electron localization factor as is done earlier \cite{pcr}. Here, it 
will be used to design a asymptotic corrected meta-GGA functional for 2D systems in the following section. Other
meta-GGA ingredient that can be used to design functionals for 2D are 
\begin{eqnarray}
W^{\sigma-KS} = \frac{\tau^{\sigma-unif}-\tau^{\sigma-KS}}{\tau^{\sigma-unif}+\tau^{\sigma-KS}};
-1\leq W^{\sigma-KS}\leq 1~.
\label{eq19}
\end{eqnarray} 
In homogeneous density limit, $W^{\sigma-KS}=0$ and in tail region $W^{\sigma-KS} \to 1$. Also, $W^{\sigma-KS}$ 
approaches $1$ as $\tau^{\sigma-unif}<<\tau^{\sigma-KS}$. Whereas, $W^{\sigma-KS} \to-1$, when $\tau^{\sigma-unif}
>>\tau^{\sigma-KS}$. The slowly varying density limit of $W^{\sigma-KS}$ is achieved by using the semi-classical 
expansion of the kinetic energy density. This gives,
\begin{equation}
W^{\sigma-KS}\approx-\frac{4}{3}q^{\sigma},
\label{eq20}
\end{equation}
where $q^{\sigma}=\frac{\nabla^2\rho^\sigma}{4k_{F^\sigma}^2\rho^\sigma}$ is the reduced laplacian density 
gradient. Another important ingredient for meta-GGA functionals is proposed by Becke \cite{b98}. Recently, 
Becke's proposed ingredient has been used to obtain the meta-GGA functional for 2D systems \cite{js}. Also,
attention has been paid to construct such functionals \cite{jps}. The 2D ingredient proposed by Becke is
\begin{equation}
Q_B^{\sigma} = \frac{1}{\tau^{\sigma-unif}}\Big[\tau^{\sigma-unif}-\tau^{\sigma-KS}+\frac{1}{4}
\frac{|\vec{\nabla}\rho^\sigma|^2}{\rho^\sigma}+\frac{1}{2}\nabla^2\rho^\sigma\Big]~.
\label{eq21}
\end{equation}
This inhomogeneity parameter can be used to diffuse or compact the exchange hole surrounding an electron. If 
the exact quadratic term i.e. the term containing inhomogeneity parameter is larger than its homogeneous 
counterpart. Then, it represents more the compact hole. If not, then the diffuse exchange hole. The $Q_B^
{\sigma}$, vanishes for uniform density and ranges from $-\infty$ to $\infty$. As near the origin, $\tau^
{\sigma-KS} \approx \tau^{\sigma-W}$. So due to the presence of the Laplacian term in $Q_B^{\sigma}$, it 
becomes $-\infty$ and at the exponential tail region it is $\infty$. It is also invariant under uniform 
coordinate scaling. A generalized coordinate transformation based $Q_B^{\sigma}$ can also be formed as is
given in \cite{jps}.

The iso-orbital indicator obtained in Eq.(\ref{eq16}) by assuming that the orbitals are real. But there is a 
possibility of non-zero current density especially in the case of applied electromagnetic field. In that case,
the current density should be added to the VW kinetic energy term and at the same time one has to ensure that 
for one electron and two electron singlet state it becomes $1$. Therefore, the following modification can be 
done to $z^\sigma$,
\begin{equation}
z^\sigma = \frac{\tilde{\tau}^{\sigma-VW}}{\tilde{\tau}^{\sigma-KS}}
\label{eq22} 
\end{equation}  
with,
\begin{eqnarray}
\tilde{\tau}^{\sigma-VW} &=& \tau^{\sigma-VW}-\frac{j_{p,\sigma}^2}{\rho^\sigma}\nonumber\\
\tilde{\tau}^{KS}_\sigma &=& \tau^{\sigma-KS}-\frac{j_{p,\sigma}^2}{\rho^\sigma},
\label{eq23}
\end{eqnarray}
where
\begin{equation}
j_{p,\sigma} = \frac{i}{2}\sum_{i=1}^{N_\sigma}\{\nabla\Psi^{*\sigma}_{i}(\vec{r})\Psi_{i}^{\sigma}(\vec{r})
-\Psi^{*\sigma}_{i}(\vec{r})\nabla\Psi_{i}^{\sigma}(\vec{r})\}~.
\label{eq24}
\end{equation}

\section{\label{sec:mgga}asymptotic corrected meta-GGA}
The asymptotic region is a physically important in 2D DFT. At asymptotic region, the exchange energy density and 
potential behave as: $\epsilon_{x\sigma} \to -\frac{1}{2r}$ and $v_{x\sigma} \to -\frac{1}{r}$ respectively (see 
appendix-A for details). Now asymptotic corrected semilocal exchange energy functionals at meta-GGA level can be 
constructed using the ingredients of the previous section. In 2D, any GGA or meta-GGA exchange energy functional 
can be written in terms of 2D-enhancement factor as,
\begin{equation}
E_{x\sigma} = \int~\rho^{\sigma}(\vec{r})\epsilon_{x\sigma}~d^2r
\label{eq25}
\end{equation}
with $\epsilon_{x\sigma}$ be the spin polarized exchange energy per particle given by
\begin{equation}
\epsilon_{x\sigma}=-A_x\epsilon^{unif}_{x\sigma}F_{x\sigma},
\label{eq26}
\end{equation}
where $\epsilon^{unif}_{x\sigma}$ is the exchange energy per particle within LDA, $A_x = \frac{4k_{F\sigma}}{3\pi}$ 
and $F_{x\sigma}$ is the spin polarized enhancement factor need to be formed from meta-GGA ingredients. Now, to 
construct $F_{x\sigma}$ having correct asymptotic behavior of exchange density and potential we have consider the 
following form,
\begin{equation}
F_{x\sigma} = \sqrt{\alpha^\sigma}.
\label{eq27}
\end{equation}
The exchange energy density can be written using this enhancement factor as, 
\begin{eqnarray}
\epsilon_{x\sigma}&=&-\frac{4(4\pi)^{1/2}}{3\pi}\rho^{\sigma}(\vec{r})^{1/2}\sqrt{\alpha^\sigma}\nonumber\\
                  &=&-\frac{4(4\pi)^{1/2}}{3\pi}\frac{1}{\sqrt{2\pi}}\frac{\Upsilon^\sigma}{\rho^\sigma}
\label{eq28}
\end{eqnarray}
with
\begin{equation} 
\Upsilon^\sigma = \sqrt{4\tau^{\sigma-KS}\rho^{\sigma}-|\nabla\rho^{\sigma}(\vec{r})|^2}~.
\label{eq29}
\end{equation}
Now to prove the asymptotic behavior of the $\epsilon_{x\sigma}$, we can plugin Eq.(\ref{eq14}) into 
Eq.(\ref{eq28}) and obtain,
\begin{equation}
\epsilon_{x\sigma} = -\frac{4(4\pi)^{1/2}}{3\pi}\sqrt{\frac{2}{\pi}}C\frac{l'}{r}~.
\label{eq30}
\end{equation}
Thus, exact asymptotic behavior is achieved, if $C=\sqrt\frac{\pi}{2}\frac{1}{2l'}$ can be included in the 
expression of enhancement factor. Now to obtain the asymptotic behavior of $v_{x\sigma}$, lets consider 
the functional derivatives of exchange energy functional,
\begin{equation}
\begin{split}
v_{x\sigma}\Psi_{i}^{\sigma} = \Big[\frac{\partial(\rho^{\sigma}\epsilon_{x\sigma})}{\partial\rho^\sigma}
-\vec{\nabla}\frac{\partial(\rho^{\sigma}\epsilon_{x\sigma})}{\partial\vec{\nabla}\rho^\sigma}\Big]
\Psi_{i}^{\sigma}-\vec{\nabla}\Big(\frac{\partial(\rho^\sigma\epsilon_\sigma)}{\partial\tau^{\sigma-KS}}\Big)
\vec{\nabla}\Psi_{i}^{\sigma}\\
-\frac{\partial(\rho^\sigma\epsilon_{x\sigma})}{\partial\tau^{\sigma-KS}}\vec\nabla^2\Psi_{i}^{\sigma}\\
=\frac{\partial(\rho^{\sigma}\epsilon_{x\sigma})}{\partial\rho^\sigma}-\vec{\nabla}\cdot\Big[\frac{\partial
(\rho^{\sigma}\epsilon_{x\sigma})}{\partial\vec{\nabla}\rho^\sigma}\Psi_{i}^{\sigma}+\frac{\partial(\rho^\sigma
\epsilon_{x\sigma})}{\partial\tau^{\sigma-KS}}\vec\nabla\Psi_{i}^{\sigma}\Big]\\
+\Big(\frac{\partial(\rho^{\sigma}\epsilon_{x\sigma})}{\partial\vec{\nabla}\rho^\sigma}\Big)\cdot\vec{\nabla}
\Psi_{i}^{\sigma}
\label{eq31}
\end{split}
\end{equation}
The terms containing Eq.(\ref{eq28}) are obtained by performing the functional derivative. This gives,
\begin{eqnarray}
\frac{\partial(\rho^{\sigma}\epsilon_{x\sigma})}{\partial\rho^\sigma} &=& -\frac{4(4\pi)^{1/2}}{3\pi}\frac{1}
{\sqrt{2\pi}}\frac{4\tau^{\sigma-KS}}{\Upsilon^\sigma} \nonumber\\
&=&-\frac{4(4\pi)^{1/2}}{3\pi}\frac{1}{\sqrt{2\pi}}\frac{\Upsilon^\sigma}{2\rho^\sigma} + \frac{\partial
(\rho^{\sigma}\epsilon_{x\sigma})}{\partial\tau^{\sigma-KS}}\frac{|\vec{\nabla}\rho^\sigma|}{2\rho^\sigma}
\frac{|\vec{\nabla}\rho^\sigma|}{2\rho^\sigma}\nonumber\\
\label{eq32}
\end{eqnarray}
and also,
\begin{eqnarray}
\frac{\partial(\rho^{\sigma}\epsilon_{x\sigma})}{\partial\vec{\nabla}\rho^\sigma}&=&\frac{4(4\pi)^{1/2}}
{3\pi}\frac{1}{\sqrt{2\pi}}\frac{|\vec{\nabla}\rho^{\sigma}|}{\Upsilon^{\sigma}}\nonumber\\
&=&-\frac{\partial(\rho^{\sigma}\epsilon_{x\sigma})}{\partial\tau^{\sigma-KS}}\frac{|\vec{\nabla}\rho^{\sigma}|}
{2\rho^{\sigma}}
\label{eq33}
\end{eqnarray}
\begin{equation}
\frac{\partial(\rho^{\sigma}\epsilon_{x\sigma})}{\partial\tau^{\sigma-KS}} = -\frac{4(4\pi)^{1/2}}{3\pi}
\frac{1}{\sqrt{2\pi}}\frac{2\rho^\sigma}{\Upsilon^{\sigma}}
\label{eq34}
\end{equation}
Now substituting Eq.(\ref{eq32}) and Eq.(\ref{eq34}) back in Eq.(\ref{eq31}) leads to
\begin{equation}
\begin{split}
v_{x\sigma}\Psi_{i}^{\sigma}=-\frac{4(4\pi)^{1/2}}{3\pi}\frac{1}{\sqrt{2\pi}}\frac{\Upsilon^\sigma}{2\rho^\sigma}
\Psi_{i}^{\sigma}-\vec{\nabla}\cdot\Big[\frac{\partial(\rho^{\sigma}\epsilon_{x\sigma})}{\partial\tau^{\sigma-KS}}\\
\Big\{-\frac{|\vec{\nabla}\rho_{\sigma}|}{2\rho^{\sigma}}\Psi_{i}^{\sigma}
+\vec{\nabla}\Psi_{i}^{\sigma}\Big\}\Big]+\frac{\partial(\rho^{\sigma}\epsilon_{x\sigma})}{\partial\tau^{\sigma-KS}}
\frac{|\vec{\nabla}\rho^{\sigma}|}{2\rho^{\sigma}}\Big\{\frac{|\vec{\nabla}\rho^{\sigma}|}{2\rho^{\sigma}}
\Psi_{i\sigma}\\-\vec{\nabla}\Psi_{i}^{\sigma}\Big\}~.
\label{eq35}
\end{split}
\end{equation}
This is the exchange potential for meta-GGA type exchange. In DFT, the highest occupied KS orbital is significant 
to explore the asymptotic property of the potential. In that case, we denote the orbital as $\Psi_{i}^{\sigma}
= \Psi^{\sigma}_H = \sqrt{\rho'}$ ($\sqrt{\rho'}$ is the asymptotic density as given in Eq.(\ref{eq14})) and 
have considered the occupation of the highest occupied label to be equal to $1$. Now, substituting $\Psi^{\sigma}_H 
= \sqrt{\rho'}$ in Eq.(\ref{eq35}), one can observe that the term inside the curly braces vanishes. Thus,
\begin{equation}
v_{x\sigma}\Psi^{\sigma}_H=-\frac{4(4\pi)^{1/2}}{3\pi}\frac{1}{\sqrt{2\pi}}\frac{\Upsilon^\sigma}{2\rho^\sigma}
\Psi^{\sigma}_H,
\label{eq36}
\end{equation}
which on imposing the asymptotic condition on $\Upsilon^\sigma$ becomes,
\begin{equation}
v_{x\sigma}\Psi^{\sigma}_H=-\frac{4(4\pi)^{1/2}}{3\pi}\frac{1}{\sqrt{2\pi}}\tilde{C}\frac{l'}{r}\Psi^{\sigma}_H~.
\label{eq37}
\end{equation}
Therefore, asymptotic behavior of $v_{x\sigma}$ will be achieved if,
\begin{equation}
\tilde{C} = \frac{\sqrt{2\pi}}{l'}
\end{equation}
In Appendix-A, we have given a scheme for obtaining all the parameters using exact or nearly exact constraints 
that should be satisfied by a 2D exchange energy functional. 

\section{\label{sec:con}conclusion}
Meta-GGA functionals are the most accurate and advanced semilocal functionals within DFT. It uses positive 
definite kinetic energy density, reduced density gradient and reduced laplacian gradient as its ingredients. 
Behavior of the kinetic energy density plays a crucial role in designing such functionals. The behavior of 
positive defined kinetic energy density for finite systems in 2D is obtained rigorously using parabolic 
quantum dot as a model system. It has been demonstrated that at origin, the KS kinetic energy density not 
only contains the VW kinetic energy density for $s$-states but $p$-states also contribute. Thus, KS-KE density 
becomes sum of the VW kinetic energies obtained from different orbitals. Similarly, at asymptotic region, 
the outermost shells contribute and the correct VW behavior of the KS kinetic energy density obtained near
the origin. Deviation of the KS kinetic energy density from its VW counterpart at asymptotic region with 
varying number of particles in a finite parabolic quantum dot is also discussed.

Next, we have obtained the behavior of several meta-GGA ingredients for the finite 2D systems i.e., parabolic 
quantum dots, which can be used to separate out different physical regions of interest. The order of limit 
problem of the KS kinetic energy density is also exemplified. As the meta-GGA ingredients are very useful in 
designing the desired enhancement factors with various properties. One such property i.e. the correct asymptotic 
behavior at the meta-GGA level for exchange energy and potential. We have obtained that property by making use 
of an wellknown meta-GGA ingredient. Then, proposed an enhancement factor which leads to a semilocal functional 
for exchange which is applicable to two dimensional quantum systems.

\section{\label{sec:ack}Acknowledgments}
The authors would like to acknowledge the financial support from the Department of Atomic Energy, Government 
of India.

\appendix
\section{\label{sec:factor}Constraints for enhancement factor}\label{ap1}
In this appendix, we will discuss the exact behavior of the enhancement factor at $\vec{r} \to 0$ and $\vec{r}
\to \infty$, using the 2D isotropic harmonic oscillator as an example. To do this, we consider the normalized 
ground state single particle wavefunction,
\begin{equation}
\Psi^{\sigma}(\vec{r}) = \frac{\alpha}{\sqrt{\pi}}\exp\Big[-\frac{\alpha^2r^2}{2}\Big]~.
\label{aeq1}
\end{equation}
Using single particle density matrix,
\begin{equation}
\varGamma_\sigma(\vec{r},\vec{r}+\vec{u}) = \sum_{\sigma=\uparrow~\textit{or}~\downarrow}\sum_{i,j=1}^{N_\sigma}
\frac{\Psi^{\sigma}_{i}(\vec{r})\Psi^{\sigma*}_{j}(\vec{r}+\vec{u})\Psi^{\sigma}_{j}(\vec{r}+\vec{u})
\Psi^{\sigma*}_{j}(\vec{r})}{\rho^\sigma(\vec{r})}~,
\label{aeq2} 
\end{equation}
the cylindrically averaged exchange hole is defined as,
\begin{equation}
\langle~h_x^{\sigma}(\vec{r}+\vec{r}+\vec{u})~\rangle = -\frac{\langle\varGamma_\sigma(\vec{r},\vec{r}+\vec{u})
\rangle}{4\rho^\sigma(\vec{r})}~.
\label{aeq3}
\end{equation}
This is further used to obtain the exchange energy density as, 
\begin{equation}
\epsilon_{x\sigma} = -\pi\int~\langle~h_x^{\sigma}(\vec{r}+\vec{r}+\vec{u})~\rangle~du.
\label{aeq4}
\end{equation}
Now using single electron wavefunction, Eq.(\ref{aeq1}), the above exchange energy density reduces to
\begin{equation}
\epsilon_{x\sigma} = -\frac{\alpha\sqrt{\pi}}{2}\exp[-\alpha^2r^2/2]I_0(\alpha^2r^2/2),
\label{aeq5}
\end{equation}
where $I_0$ is the zeroth order modified Bessel function. We are interested in two regions, namely $\vec{r}
\to 0$ and $\vec{r}\to \infty$. For $\vec{r}\to 0$, $I_0(0)=1$ and for $\vec{r}\to\infty$ region, $I_0(\infty)
= \frac{\exp(\alpha^2r^2/2)}{\alpha r \sqrt{\pi}}$. Thus,
\begin{eqnarray}
\epsilon_{x\sigma}(\vec{r}\to 0) &=& -\frac{\alpha\sqrt{\pi}}{2}\nonumber\\
\epsilon_{x\sigma}(\vec{r}\to \infty)&=&-\frac{1}{2r}.
\label{aeq6}
\end{eqnarray}
The above results are used to obtain the asymptotically correct exchange energy functional using the meta-GGA 
ingredients. Having established the desired behavior of the exchange energy density, we now derive the $\vec{r} 
\to 0$ and $\vec{r} \to \infty$ behavior of the enhancement factor 
\begin{eqnarray}
F_{x\sigma} &=& \frac{\epsilon_{x\sigma}(\vec{r})}{\epsilon_{x\sigma}^{unif}(\vec{r})}\nonumber\\
&=&\frac{3\pi^{3/2}}{16}I_0\Big(\frac{\alpha^2r^2}{2}\Big)~.
\label{aeq7}
\end{eqnarray}
Thus,
\begin{eqnarray}
F_{x\sigma}(\vec{r}\to 0) &=& \frac{3\pi^{3/2}}{16}\nonumber\\
F_{x\sigma}(\vec{r}\to \infty)&=& \frac{3\pi}{16}\frac{\exp\Big(\frac{\alpha^2r^2}{2}\Big)}{\alpha r}~.
\label{aeq8}
\end{eqnarray}
For slowly varying density, by making use of the semi-classical approximation of the kinetic energy density, 
one can obtain the form of $\alpha^\sigma$ that has been used to construct the enhancement factor. This gives
\begin{equation}
\alpha^{\sigma} = 1 - 2p^{\sigma} + \frac{8}{3}q^{\sigma}~.
\label{aeq9}
\end{equation}
As the enhancement factor has the form
\begin{equation}
F_{x\sigma}\to \Big[1 - 2p^{\sigma} + \frac{8}{3}q^{\sigma}\Big]^{\frac{1}{2}}\approx 1 - p^{\sigma} + 
\frac{4}{3}q^{\sigma}~.
\label{aeq10}
\end{equation}
The exchange energy becomes,
\begin{equation}
E_{x\sigma} = \int~d^2r~\rho^{\sigma}\epsilon_{x\sigma}^{unif}\Big[1 - p^{\sigma} + \frac{4}{3}q^{\sigma}\Big]~.
\label{aeq11}
\end{equation}
On performing integration by parts, the reduced laplacian gradient can be transformed into reduced density gradient. 
Thus we obtain,
\begin{equation}
E_{x\sigma} = \int~d^2r~\rho^{\sigma}\epsilon_{x\sigma}^{unif}\Big[1+\mu^{SGL}p^{\sigma}\Big],
\label{aeq12}
\end{equation} 
where $\mu^{SGL}$ is the coefficient of enhancement factor for slowly varying density obtained through the 
small density gradient approximation of enhancement factor \cite{jps}. The values of enhancement factor 
obtained in different regions are as follows:

(i) Near origin or $\vec{r}\to 0$
\begin{equation}
F_{x\sigma} \to \frac{3\pi^{3/2}}{16};
\label{aeq13}
\end{equation} 

(ii) Asymptotic region or tail region or $\vec{r}\to \infty$
\begin{equation}
F_{x\sigma} \to \sqrt{\alpha^{\sigma}};
\label{aeq14}
\end{equation}

(iii) Slowly varying density limit
\begin{equation}
F_{x\sigma} \to 1 + \mu^{SGL}p^{\sigma};
\label{aeq15}
\end{equation}

(iv) $F_{x\sigma} \geq 0$ everywhere.

All these exact or nearly exact constraints can be used to design the meta-GGA ingredients based exchange energy 
functionals. One such functional, we have obtained above. Not only that, this is also useful to obtain other 
meta-GGA ingredient based enhancement factors.

\end{document}